\input harvmac
\input amssym.def
\input amssym.tex
\input epsf.tex

\font\male=cmr9
\font\tfont=cmbx12 scaled\magstep1

\def\newsubsec#1{\global\advance\subsecno by1\message{(\secsym\the\subsecno.
#1)} \ifnum\lastpenalty>9000\else\bigbreak\fi
\noindent{\bf\secsym\the\subsecno. #1}\writetoca{\string\quad
{\secsym\the\subsecno.} {#1}}}

\def\newsubsubsec#1{\global\advance\subsubsecno
by1\message{(\secsym\the\subsecno.\the\subsubsecno.
#1)} \ifnum\lastpenalty>9000\else\bigbreak\fi
\noindent{\bf\secsym\the\subsecno.\the\subsubsecno. #1}\writetoca{\string\quad
{\secsym\the\subsecno.\the\subsubsecno.} {#1}}}

\hfuzz=15pt

\def\nt{\noindent}
\def\nl{\hfil\break}

\newcount\figno
\figno=0
\def\fig#1#2#3{
\par\begingroup\parindent=0pt\leftskip=1cm\rightskip=1cm\parindent=0pt
\baselineskip=11pt
\global\advance\figno by 1
\midinsert
\epsfxsize=#3
\centerline{\epsfbox{#2}}
\vskip 12pt
#1\par
\endinsert\endgroup\par}
\def\figlabel#1{\xdef#1{\the\figno}}
\def\encadremath#1{\vbox{\hrule\hbox{\vrule\kern8pt\vbox{\kern8pt
\hbox{$\displaystyle #1$}\kern8pt}
\kern8pt\vrule}\hrule}}


\def\l{\lambda}

\def\({\left(} \def\){\right)} 
\def\k{{\kappa}}   \def\lra{\longrightarrow}
 
\def\cp{{\cal P}} \def\tV{\widetilde V}  \def\bac{{C\kern-5.2pt I}}
\def\h#1{\widehat{#1}} \def\vv{{\tilde v}_0} \def\th{{\tilde h}}
\def\bbq{{\rm Q}\kern-7pt {\rm I}\ }
\def\bbc{{C\kern-6.5pt I}}
\def\bbn{I\!\!N}
\def\bbz{Z\!\!\!Z}

 \def\eps{\epsilon}
\def\L{\Lambda}  \def\tp{{\tilde p}} \def\tq{{\tilde q}}
\def\hW{{\widehat W}} \def\hR{{\widehat R}} \def\hB{{\widehat B}}
\def\hQ{{\widehat Q}} \def\hS{{\widehat S}} \def\hC{{\widehat C}}
\def\hc{{\hat c}}   \def\otw{{\textstyle{1\over 12}}}
\def\pot{{\textstyle{p\over 2}}}\def\qot{{\textstyle{q\over 2}}}
\def\pott{{\textstyle{p+1\over 2}}}

\def\otf{{\textstyle{1\over 24}}}
\def\osi{{\textstyle{1\over 16}}}\def\osic{{\textstyle{c\over 16}}}
\def\d{\delta} \def\half{{\textstyle{1\over2}}}
\def\oei{{\textstyle{1\over 8}}} \def\nnu{{\textstyle{1\over \nu}}}
\def\qtr{{\textstyle{1\over 4}}}\def\qtrn{{\textstyle{1\over 4\nu}}}
\def\thf{{\textstyle{3\over2}}}
\def\a{\alpha}\def\b{\beta}

\def\cp{{\cal P}}

\lref\BPZ{A.A. Belavin, A.M. Polyakov and A.B. Zamolodchikov,
Nucl. Phys. {\bf 241} (9184) 333.}
\lref\Fel{G. Felder, Nucl. Phys. {\bf B317} (1989) 215,
Erratum-ibid. {\bf B324} (1989) 548.}
\lref\Seg{G. Segal, Comm. Math. Phys. {\bf 80} (1981) 301.}
\lref\CDK{N. Chair, V.K. Dobrev and H. Kanno, Phys. Lett. {\bf 283B} (1992)
194-202.}
\lref\FST{P. Furlan, G.M. Sotkov and I.T. Todorov,
J. Math. Phys. {\bf 28} (1987) 1598.}
\lref\BDIZ{M. Bauer, P. Di Francesco, C. Itzykson and J.-B. Zuber,
Phys. Lett. {\bf 260B} (1991) 323-326;
Nucl. Phys. {\bf B362} (1991) 515-562.}
\lref\Kent{A. Kent, Phys. Lett. {\bf 273B} (1991) 56-62;\
ibid. {\bf 278B} (1992) 443-448.}
\lref\GaPe{A.Ch. Ganchev and V.B. Petkova,
Phys. Lett. {\bf 293B} (1992) 56-66; Phys. Lett. {\bf 318B} (1993) 77-84.}

\lref\KK{V.G. Kac and D. Kazhdan, ``{\it Structure of
representations with highest weight of infinite--dimensional Lie
algebras}'', Adv. Math. {\bf 34} 97--108 (1979).}

\lref\Lan{R.P. Langlands,
"On unitary representations of the Virasoro algebra", in:
Infinite-dimensional Lie algebras and their applications,
(World Scientific, 1988) 141-159.}

\lref\BeSaa{L. Benoit and Y. Saint-Aubin,
Phys. Lett. {\bf 215B} (1988) 517;
Lett. Math. Phys. {\bf 23} (1991) 117-120;
Int. J. Mod. Phys. {A7} (1992) 3023-3033.}
\lref\GFV{I.M. Gel'fand and D.B. Fuchs,
Funkts. Anal. Prilozh. {\bf 2} (4) (1968) 92-93;\nl
M.A. Virasoro, Phys. Rev. {\bf D1} (1970) 2933-2936.}
\lref\NS{A. Neveue and J.H. Schwarz, Nucl. Phys. {\bf B31}
(1971) 86-112.}
\lref\Ram{P. Ramond, Phys. Rev. {\bf D3} (1971) 2415-2418.}
\lref\Kac{V.G. Kac, in: Proc. of ICM, Helsinki (1978) 299-304;\
Lecture Notes in Physics, Vol. 94 (1979) pp. 441-445.}
\lref\FQS{D. Friedan, Z. Qiu and S. Shenker,
Phys. Rev. Lett. {\bf 52} (1984) 1575-1578,\
Phys. Lett. {\bf 151B} (1985) 37-43, Comm. Math. Phys.
{\bf 107} (1986) 535-542.}
\lref\FFb{B.L. Feigin  and D.B. Fuchs,
Funkts. Anal. Prilozh. {\bf 17} (3) (1983) 91-92;\
English translation: Funct. Anal. Appl. {\bf 17} (1983) 241-242.}
\lref\FFa{B.L. Feigin  and D.B. Fuchs,
Funkts. Anal. Prilozh. {\bf 16} (2) (1982) 47-63;\
English translation: Funct. Anal. Appl. {\bf 16} (1982) 114-126.}
\lref\FFc{B.L. Feigin  and D.B. Fuchs,
Lecture Notes in Math. vol. 1060 (1984) pp. 230-245.}
\lref\RC{A. Rocha-Caridi, in: 'Vertex Operators in
Mathematics and Physics', eds. J. Lepowsky, S. Mandelstam
and I. Singer (Springer, New York, 1985) pp. 451-473.}

\lref\Kaca{V.G. Kac, Lecture Notes in Math. vol. 933 (1982) pp. 117-126.}
\lref\RCW{A. Rocha-Caridi and N.R. Wallach, Math. Z. {\bf 185}
(1984) 1-21.}
\lref\KW{V.G. Kac and M. Wakimoto,
Lecture Notes in Physics, Vol. 261 (1986) pp. 345-371.}
\lref\IZ{C. Itzykson and J.-B. Zuber, Nucl. Phys. {\bf B275}
[FS17] (1986) 580-616.}
\lref\Doo{V.K. Dobrev, ``{\it Multiplet classification of the
indecomposable highest weight modules over affine Lie algebras
and invariant differential operators: the $A^{(1)}_\ell$
example}'', Talk at the Conference on Algebraic Geometry and
Integrable Systems, Oberwolfach, July 1984 and ICTP, Trieste,
preprint, IC/85/9 (1985).}
\lref\GKO{P. Goddard, A. Kent and D. Olive, Comm. Math. Phys.
{\bf 103} (1986) 105-119.}
\lref\Dov{V.K. Dobrev, in: Proceedings of the XIII
International Conference on Differential-Geometric Methods in
Theoretical Physics, Shumen (1984), eds. H.D. Doebner and T.D.
Palev, (World Sci., Singapore, 1986) pp. 348-370.}
\lref\Dons{V.K. Dobrev, Lett. Math. Phys. {\bf 11} (1986) 225-234.}
\lref\Dosr{V.K. Dobrev, Lectures at the Srni Winter School (January
1986), Suppl. Rendiconti Circolo Matematici di Palermo,
Serie II, Numero 14 (1987) 25-42.}


\centerline{{\tfont Representations and characters of the}}
\centerline{{\tfont Virasoro algebra and N=1
super-Virasoro algebras} \foot{This is a slightly extended version
of an Encyclopedia entry.}}

\vskip 0.5cm

\centerline{{\bf V.K. Dobrev}}
\centerline{\male Institute for Nuclear Research and Nuclear Energy}
\centerline{\male Bulgarian Academy of Sciences, Sofia, Bulgaria}

\vskip 0.5cm

\centerline{{\bf Abstract}}
\midinsert\narrower{\male
We present the list of irreducible (generalized) highest weight modules
over the Virasoro algebra and N=1 super-Virasoro algebras obtained as factor-modules
of  (generalized) Verma modules. We present also the character formulae
of all these modules and single out the unitary irreducible ones.
Most formulas are valid for the three algebras under consideration,
the different cases being distinguished by two parameters.
 }\endinsert

\vskip 0.5cm

\newsec{Representation theory}

\nt
The ~{\it Virasoro algebra}\ \GFV\ \ $\hW$ ~is a complex
Lie algebra with basis
~$\hc, L_n\,$, $n\in\bbz$ and Lie brackets:
 \eqna\vir
$$\eqalignno{
[L_m,L_n] ~~=&~~ (m-n)\, L_{m+n} ~+~ \d_{m,-n}\, \otw\,\left( m^3-m \right)\,
  \hc &\vir a\cr
[\hc,L_n]  ~~=&~~ 0 &\vir b\cr }$$

The ~{\it Neveu-Schwarz superalgebra}\ \NS\ \ $\hS$ ~is a complex Lie
superalgebra with basis
~$\hc, L_n\,, J_\a$, $n\in\bbz$, $\a\in\bbz + \half\,$,
and Lie (super-)brackets:
 \eqna\virn
$$\eqalignno{
[L_m,L_n] ~~=&~~ (m-n)\ L_{m+n} ~+~ \d_{m,-n}\ \oei\,\left( m^3-m \right)\
\hc &\virn a\cr
[J_\a,J_\b]_+ ~~=&~~ 2\, L_{\a+\b} ~+~ \d_{\a,-\b}\ \half\,
\left( \a^2-\qtr \right)\  \hc
&\virn b\cr
[L_m,J_\a] ~~=&~~ (\half m-\a)\, J_{m+\a}  &\virn c\cr
[\hc,L_n]  ~~=&~~ 0 \ , \quad [\hc,J_\a]  ~~=~~ 0
 &\virn d\cr }$$

The ~{\it Ramond superalgebra}\ \Ram\ \ $\hR$ ~is a complex Lie
superalgebra with basis $\hc, L_n$, $n\in \bbz$, $J_\alpha\,$,
$\alpha\in \bbz$ and Lie brackets given again by \virn{}.

The Neveu-Schwarz and Ramond superalgebras are also called
~$N=1$~ {\it super-Virasoro algebras}, since they can be viewed
as $N=1$ supersymmetry extensions of the Virasoro algebra.

Further, $\hQ$ will denote $\hW$, $\hS$ or $\hR$ when a statement
holds for all three algebras. The elements $\hc,L_n$ are even and $J_\a$
are odd. The grading of $\hQ$ is given by:
\eqn\gra{ \deg\hc ~=~ 0, \quad \deg L_n ~=~ n , \quad \deg J_\a ~=~ \a }
We have the obvious decomposition:
\eqn\dec{ \hQ ~~=~~ \hQ_+ \ \oplus\ \hQ_0 \ \oplus\ \hQ_- }
where
\eqna\dcc
$$\eqalignno{
\hQ_+  ~~=&~~ {\rm c.l.s.} \{\ X\in \hQ ~\vert~ \deg X >0 \} &\dcc a\cr
\hQ_0  ~~=&~~ {\rm c.l.s.} \{\ X\in \hQ ~\vert~ \deg X =0 \} &\dcc b\cr
\hQ_-  ~~=&~~ {\rm c.l.s.} \{\ X\in \hQ ~\vert~ \deg X <0 \} &\dcc c\cr}$$
(c.l.s. means `complex linear span').
Thus, $\hW_0$ and $\hS_0$ are spanned by $\hc$ and $L_0\,$, while
$\hR_0$ is spanned by $\hc$, $L_0$ and $J_0\,$.
Further, we note that $\hW_\pm$ is generated by $L_{\pm 1}$ and $L_{\pm 2}\,$,
$\hS_\pm$ is generated by $J_{\pm \half}$ and $J_{\pm \thf}\,$,
$\hR_\pm$ is generated by $L_{\pm 1}$ and $J_{\pm 1}\,$.

Next we note that the Cartan subalgebra $\hC$
of $\hQ$ is spanned by $\hc$ and $L_0\,$. We note that $\hC = \hQ_0$
for $\hW$ and $\hS$. In the case of $\hR$ the generator $J_0$ is
not in $\hC$ since it is odd and can not diagonalize $\hR_\pm$ - having an
anti-bracket with the odd generators. Because of this peculiarity
we shall use generalized highest weight modules, which will be relevant
only in the $\hR$ case, since in the other two cases they will coincide
with ordinary highest weight modules.

A {\it generalized highest weight module}\ over $\h{Q}$ is characterized by
its highest weight $\Lambda \in \hC^*$ and {\it generalized highest weight
vector}\ ${\tilde v}_0\,$, which is a finite-dimensional vector space,
so that:
\eqn\hwm{\eqalign{
 X\,v ~=&~ 0\ , \quad X\in\hQ_+\,,\ v\in\vv \cr
X\,v ~=&~ \L(X)\,v\ , \quad X\in\hC,\ v\in\vv }}
We define the generalized highest weight vector ${\tilde v}_0$ for our
three algebras by:
\eqna\ghwm
$$\eqalignno{  {\tilde v}_0 ~=&~ c.l.s.\,\{\, v_0\,\} \,,
\quad {\rm for}\ \ \hW,\hS
&\ghwm a\cr
{\tilde v}_0 ~=&~ c.l.s.\, \{\, v_0\,, J_0\,v_0\,\}  \,,
\quad {\rm for}\ \ \hR
&\ghwm b}$$
where $v_0$ fulfills the conditions \hwm, i.e., it is
a usual highest weight vector. We denote:
\eqn\hww{ \L(L_0) ~=~ h , \quad \L(\hc) ~=~ c }

Further we shall need also generalized Verma modules over
$\h{Q}$. For  this we introduce bases in the universal enveloping
algebras ~$U(\hQ_\pm)$, $U(\hQ_0)$~ as follows.
\eqna\bass
$$\eqalignno{ &J_{\a_1} \dots J_{\a_k}\ L_{n_1} \dots L_{n_\ell} \,, \quad
0<\a_1 < \cdots < \a_k  \,, \quad 0<n_1 \leq \cdots \leq n_\ell \,,
\quad {\rm for} ~\hQ_+\quad &\bass a\cr
&J_{\a_1} \dots J_{\a_k}\ L_{n_1} \dots L_{n_\ell} \,, \quad
\a_1 < \cdots < \a_k <0  \,, \quad n_1 \leq \cdots \leq n_\ell <0\,,
\quad {\rm for} ~\hQ_-\quad &\bass b\cr
& \hc^k \ L_0^\ell \ J_0^\k \,, \quad
k,\ell\in\bbz_+ \,, \quad \k =0  \,,\quad {\rm for} ~\hW_0 \,,\hS_0
\,, \quad \k =0,1  \,,\quad {\rm for} ~\hR_0
&\bass c\cr }$$
Naturally, the odd elements appear at most in first degree,
since one has (cf. \virn{b}):
\eqna\rrel
$$\eqalignno{
J_\a^2 ~=&~  L_{2\a} \,, \qquad \a\neq 0 &\rrel a\cr
J_0^2  ~=&~  L_0 ~-~ \osi\, \hc \ ,\quad {\rm for} \ \ \hR_0
&\rrel a\cr}$$

For further reference we say that an element ~$u$~ of $U(\h{Q})$
is ~{\it homogeneous}~ if it is an eigenvector of the generator $L_0\,$,
i.e., if ~$[L_0,u] ~=~ \l_u\,u$. Then we define the ~{\it level}~
of a homogeneous element ~$u$~ of $U(\h{Q})$ as ~$-\l_u\,$.
The basis elements in \bass{a,b} are homogeneous of level
~$-(\a_1+\cdots + \a_k + n_1 + \cdots + n_\ell)$, (which
is negative, positive, resp., for $U(\hQ_+)$, $U(\hQ_-)$, resp.),
while those in \bass{c} are homogeneous of level zero.

A generalized Verma module $V^\Lambda = V^{h,c}$ is an induced GHWM
with highest weight $\Lambda$ such that \eqn\gver{ V^\Lambda\ \cong\
U(\h{Q})\, \otimes_{U(\hB)}\, {\tilde v}_0 \ \cong\ \cases{
U(\h{Q}_-)\, \otimes\, {\tilde v}_0 &~for $\hW_0 \,,\hS_0$\cr
U(\h{Q}_-)\, \otimes_{\bac\, J_0}\, {\tilde v}_0 &~for $\hR_0$ }}
where $\hB = \hQ_+ \oplus \hQ_0\,$. Below for simplicity we shall
write ~$U(\hQ_-)\ \vv$~ omitting the tensor sign. We denote by
~$L_\L ~=~ L^{h,c}$~ the irreducible factor-module ~$V^\L/I^\L$,
where ~$I^\L$~ is the maximal proper submodule of ~$V^\L$. Then
every irreducible GHWM over $\hQ$ is isomorphic to some $L_\L$.

From now on most formulas will be valid for the three algebras,
the different cases being distinguished by two parameters:
\eqn\param{\mu ~\equiv~ \cases{ 0 & for ~$\hW,\hS$\cr
\half & for ~$\hR$ } ~,\qquad
\nu ~\equiv~ \cases{ 1 & for ~$\hW$\cr
2 & for ~$\hS,\hR$ }}

It is known (cf. \Kac\ for $\hW$, $\hS$, \FQS\ for $\hR$)
that the generalized Verma module $V^\L$ is reducible iff $h$ and $c$ are
related as follows:
\nt
{\it either}
\eqna\redv
$$\eqalignno{
&h ~=~ h_{(m,n)} ~=~ h_0 \ +\ \qtr (\a_+ m + \a_- n)^2 + \oei\mu
&\redv a\cr}$$
where
$$\eqalignno{&\eqalign{
m,n \in & {1\over \nu}\,\bbn \,, \quad m-n \in\bbz +\mu\cr
h_0 ~=&~ \cases{\otf\, (c-1) & for ~$\hW$\cr
\osi (c-1) & for ~$\hS,\hR$ }\cr
\a_\pm  ~\equiv&~ \cases{{1\over \sqrt{24}}\,
\left( \sqrt{1-c} \pm \sqrt{25-c} \right) & for ~$\hW$\cr
\half \left( \sqrt{1-c} \pm \sqrt{9-c} \right) & for ~$\hS,\hR$ }}
&\redv b\cr}$$
{\it or}
\eqn\redr{ h = {c\over 16}\ \ \ \  {\rm for} \ \ \ \  \h{R} }

For further use we also introduce the notation:
\eqn\nnot{c_0 ~\equiv~ \cases{ 25 & for ~$\hW$\cr
 9 & for ~$\hS,\hR$}}

We first comment on the cases from \redv{}.
We know that the reducible generalized Verma module $V=V^{h,c}$,
$h=h_{(m,n)}\,$, contains a proper submodule isomorphic to the
generalized Verma module $V'=V^{h+\nu mn,c}$.
In other words there exists a non--trivial embedding map
between $V'$ and $V$.
In this situation we shall use the following pictorial representation:
\eqn\emb{ V^{h,c} ~\lra~ V^{h+\nu mn,c} ~, \qquad h=h_{(m,n)} }
These embedding maps are realized by the so called singular vectors.
A {\it singular vector} $v_s\in V$ is such that
$v_s \not= {\tilde v}_0$
and $v_s$ has the property of the highest weight vector
${\tilde v}'_0$ of $V'$.
More than this $v_s$ can be expressed, as an
element of $U(\h{Q}_-){\tilde v}_0$ by
\eqn\svv{  v_s = {\cal P}(\h{Q}_-){\tilde v}_0\,, \quad
V' \cong U(\h{Q}_-){\tilde v}'_0 \cong U(\h{Q}_-)
{\cal P}(\h{Q}_-){\tilde v}_0 ~, }
where $\cp(\h{Q}_-)$ is a homogeneous polynomial in $U(\h{Q}_-)$ of level
~$\nu mn$.

Explicit examples of singular vectors exist for low levels
and/or via special constructions, cf. \BPZ,
\FST, \BDIZ, \Kent, \GaPe, (all for $\hW$),
and also for the cases ~$h ~=~ h_{(m,1)},h_{(1,n)}$, cf.
\BeSaa.

Next we comment on \redr. First
we note the following decomposition in the $\hR$ case which
holds for all generalized Verma modules:
\eqn\dec{ \eqalign{
V^\Lambda ~=&~ V^\Lambda_0 \otimes V^\Lambda_1\cr
V^\Lambda_k ~=&~ U(\hQ_-)\ J_0^k\ v_0 }}
Further we consider the action of  $\hR$ on $V^\L$.
The action of $\hR_-$ and $\hC$ preserves each $V^\Lambda_k\,$.
Further, we consider separately the cases $h\neq \osi c$
and $h = \osi c$~: \nl {\bf 1.}\ \ For $h\neq \osi c$
the action of $J_0$ is not preserving $V^\Lambda_k\,$.
Indeed, it mixes the elements of $\vv$~:
\eqn\actt{ J_0\, (a v_0 + b J_0 v_0) ~=~ b (h-\osi c) v_0 + a J_0 v_0 }
Furthermore shifting $J_0$ to the right until it reaches
$\vv$ we have to use:
\eqn\act{ \eqalign{
J_0\,L_n ~=&~ L_n\,J_0 - \half n J_n \,, \quad  n<0 \cr
J_0\,J_\a ~=&~ -J_\a\,J_0 + 2 L_\a \,, \quad \a <0 }}
Finally, the action of $\hR_+$ is not preserving $V^\Lambda_k\,$, since
shifting its elements to the right one produces also $J_0\,$, since:
\eqn\actp{ \eqalign{
L_n\,J_{-n} ~=~ J_{-n}\,L_n + \thf n J_0 \,, \quad  n>0 \cr
J_n\,L_{-n} ~=~ L_{-n}\,J_n + \thf n J_0 \,, \quad  n>0 \cr }}
{\bf 2.}\ \ In the case $h=\osi c$ the subspace $V^\Lambda_1$
is preserved by the action of $\hR$. For this it is enough to
note that:\ $J_0 (J_0 v_0) ~=~  (h-\osi c) v_0 ~=~ 0$.
Thus, the action of $J_0$ and, consecutively, of $\hR_+$
on $V^\Lambda_1$ can not produce elements of $V^\Lambda_0\,$.

Thus, for $h=\osi c$ the generalized Verma module $V = V^{c/16,c}$
over $\hR$ is reducible. It contains a proper submodule $V^{c/16,c}_1$
which is isomorphic to an ordinary Verma module
${\tilde V} = {\tilde V}^{c/16,c}$
with the same highest weight and highest weight vector
$v'_0 = J_0 v_0$ with the additional property: $J_0\,v'_0 = 0$.
The factor-module $V^{c/16,c}/V^{c/16,c}_1$
is also isomorphic to ${\tilde V}$ since it has the same highest
weight and its highest weight vector $v''_0 = v_0$ also fulfils
the property to be annihilated by $J_0$ - indeed, we have
$J_0\,v''_0 = 0$ since $J_0\,v_0\in V^{c/16,c}_1\,$.
For this reason further in the case  $h=\osi c$ we shall
consider $\tV$ instead of $V^{c/16,c}$ for  $\hR$.

It is also possible that \redv{} and \redr\ hold simultaneously.
In this case ${\tilde V}$ is further reducible and everything we said
for cases from \redv{} applies also to this case. Furthermore
combining \redv{} and \redr\ we have:
\eqna\redvr
$$\eqalignno{ h ~=&~ \osi c ~=~ h_{(m,n)} ~=~
\osi (c-1) + {1\over 4}(m\alpha_+ + na_-)^2 + \osi
\ \ , \ \  {\rm or}, &\redvr a\cr
&   m\alpha_+ + n\alpha_-  = 0\ \ , \ \  {\rm or} , &\redvr b\cr
&  -\alpha_-/\alpha_+ = {m\over n} = {2m\over 2n} = {p\over q}
&\redvr c}$$
Combining this with the requirement that  $m-n \in \bbz + {1\over 2}$
or ~$2m - 2n \in 2\bbz +1$, we see that one of the two numbers
$2m,2n$ must be odd and the other even, and then one of the two numbers
$p,q$ must be odd and the other even. Thus, we see that the only
possibility in this case is:
\eqn\rdvr{ m ~=~ \half p \,, \quad n ~=~ \half q \,, \quad pq\in 2\bbn.}

\newsec{Multiplet classification of the reducible (generalized)
Verma modules}

\nt
Here we present the multiplet classification of the reducible
(generalized) Verma modules
over $\hW$ \FFb\ (following \Dov) and over $\hS$, $\hR$ \Dons.
There are five types in each case which will
be denoted (following \Dons) $N^0, N^1_+, N^1_-, N^2_+, N^2_-$
They are shown in Table 1. (Note that in \FFb,\Dov\ the notation was
~$II,III_\pm\,,III^0_\pm\,,(III^{00}_\pm)$, resp.)

The type $N^0$ occurs when the ratio $\a_-/\a_+$ is not a real
rational number.

For all other types the ratio $\a_-/\a_+$ is a real
rational number and ~{\it either}\ $c\leq 1$ ~{\it or}\ $c\geq c_0\,$.

For $c < 1$ we have type $N^1_-$ and subtypes $N^{21}_-, N^{22}_-$
of type $N^2_-\,$. In this case the ratio $\a_-/\a_+$
is negative so we set: ~$\a_-/\a_+ ~=~ -p/q$, $p,q\in\bbn$, $pXq$ (the
latter means that $p,q$ have no common divisors), and then we have:
\eqn\ccc{\eqalign{
c ~=~ c^-_{p,q} ~=&~  \half (c_0+1) - \qtr (c_0-1)\,
\left({p\over q} + {q\over p}\right) ~=~
1- (10-4\nu){(p-q)^2\over pq} ~=\cr
 ~~=&~~
\cases{13 -6 \left({p\over q} + {q\over p}\right)
& for ~$\hW$\cr
5 -2 \left({p\over q} + {q\over p}\right)
& for ~$\hS,\hR$ } \,, \qquad (c<1) }}
Substituting $c\to c^-_{p,q}$ in \redv{a} we get:
\eqn\redh{ h ~=~ h^-_{(m,n)} ~=~  {1\over 4\nu pq}
[\nu^2(pn - qm)^2 - (p-q)^2] + \oei\mu }

For $c=1$ we have subtype $N^{23}_-$ of type $N^2_-\,$.

For $c > c_0$ we have type $N^1_+$ and subtypes
$N^{21}_+, N^{22}_+$ of type $N^2_+\,$.
Here the ratio $\a_-/\a_+$
is positive so we set: ~$\a_-/\a_+ ~=~ p/q$, $p,q\in\bbn$, $pXq$,
and then we have:
\eqn\ccc{\eqalign{
c ~=~ c^+_{p,q} ~=&~ \half (c_0+1) + \qtr (c_0-1)\,
\left({p\over q} + {q\over p}\right) ~=~
1 + (10-4\nu){(p-q)^2\over pq} ~=\cr
 ~~=&~~
\cases{13 +6 \left({p\over q} + {q\over p}\right)
& for ~$\hW$\cr
5 +2 \left({p\over q} + {q\over p}\right)
& for ~$\hS,\hR$ } \,, \qquad (c>c_0) }}
Substituting $c\to c^+_{p,q}$ in \redv{a} we get:
\eqn\redhp{ h ~=~ h^+_{(m,n)} ~=~
{1\over 4\nu pq} [(p+q)^2 - \nu^2 (pn - qm)^2]
+ \oei\mu
~=~ {1\over \nu} + \qtr\mu - h^-_{(m,n)} }

For $c = c_0$ we have subtype $N^{23}_+$ of type $N^{2}_+\,$.

The explicit parametrization of all types and subtypes
is given in Table 2
(for $\hW$ see \FFb\ and Propositions 1.1-1.5, formulae (6),(13),(16)
of \Dov, for $\hS,\hR$ see Propositions 1-4, formulae (6),(11),(14)
of \Dons). Note that in each case the (sub)types $N_-$ with
$c\leq 1$ have the same parametrization as the corresponding $N_+$ with
$c\to c_0+1-c\geq c_0$. Further, we note that the parametrization
of subtypes $N^{21}_\pm$ is obtained from the parametrization of $N^{1}_\pm$
by formally setting $n=\tq$ and replacing the condition $p<q$ by $p\neq q$.
Next, the parametrization
of subtypes $N^{22}_\pm$ is obtained from the parametrization of $N^{21}_\pm$
by formally setting $m=\tp$ or $m=0$. Finally, the parametrization
of subtypes $N^{23}_\pm$ is obtained from the parametrization of $N^{22}_\pm$
by formally setting $p=q$ (which forces $p=q=1$).

Further we give the explicit parametrization of the (generalized)
Verma modules in the different multiplets. The cases  $N^{0}$
are simple and all info about then is already present in Tables 1,2.

We start with type $N^{1}_-\,$.
For fixed parameters $p,q,m,n$ (cf. Table 2)
the (generalized) Verma modules of this type form a multiplet
represented by the corresponding commutative diagram
of Table~1. The modules of this multiplet are divided in
four infinite groups which are given as follows
(cf. \Dov, formula (10), \Dons, formula (10)):
\eqna\param
$$\eqalignno{
&V_{0k} ~=~ V^{h_{0k}\,,\,c} \,,\quad k\in\bbz_+ &\param a\cr
h_{0k} ~=&~ h_{00}+\nu k(\tp\tq k + \tq m -\tp n)
~=~ h^-_{(2k\tp+m,n)} ~=~
h^-_{(\tp-m, 2k\tq+\tq-n)} &\param {a'}\cr
&V_{1k} ~=~ V^{h_{1k}\,,\,c} \,,\quad k\in\bbz_+ &\param b\cr
h_{1k} ~=&~ h_{00} +\nu k(\tp\tq k - \tq m +\tp n)
~=~ h^-_{(m,2k\tq+n)} ~=~
h^-_{(2k\tp+\tp-m, \tq-n)} &\param {b'}\cr
&V'_{0k} ~=~ V^{h'_{0k}\,,\,c} \,,\quad k\in\bbz_+ &\param c\cr
h'_{0k} ~=&~ h_{00}+\nu (\tq k +n)(\tp k + m ) ~=~
h^-_{(2k\tp+\tp+m,\tq-n)} ~=~
h^-_{(\tp-m, 2k\tq+\tq+n)} &\param {c'}\cr
&V'_{1k} ~=~ V^{h'_{1k}\,,\,c} \,,\quad k\in\bbz_+ &\param d\cr
h'_{1k} ~=&~ h_{00} +\nu (\tq k +\tq -n)(\tp k + \tp -m )
~=~ h^-_{(m,2k\tq+2\tq-n)} ~=~
h^-_{(2k\tp+2\tp-m,n)} \qquad &\param {d'}\cr
}$$
where ~$c = c^-_{p,q}\,$, and we note that ~$h_{00} =h_{10}\,$.

The (generalized) Verma modules of type $N^{1}_+$ for fixed parameters
$p,q,m,n$
also form a multiplet represented by the corresponding commutative
diagram of Table~1. The modules of this multiplet are also divided in
four infinite groups which we denote by $V^+_{0k}\,$, $V^{'+}_{0k}\,$,
$V^+_{1k}\,$, $V^{'+}_{1k}\,$, which are given by \param{}
with the changes ${\tilde p} \to -{\tilde p}, m\to -m$
(or ${\tilde q} \to -{\tilde q}, n \to -n$), ~$h\to h^+_{(m,n)}$,
~$c\to c^+_{p,q}\,$.

We continue with subtype $N^{21}_-\,$.
For fixed parameters $p,q,m$ (cf. Table 2)
the (generalized) Verma modules of this type form a multiplet
represented by the corresponding  diagram
of Table~1. The modules of this multiplet are divided in
two infinite groups whose parametrization is obtained from
those of type $N^{1}_-$ by setting $n=\tq$~:
\eqna\paran
$$\eqalignno{ V^0_{0k} ~=&~ \(V_{0k}\)_{\vert_{n=\tq}} ~=~
\(V'_{0,k-1}\)_{\vert_{n=\tq}\,,\,k\geq 1} ~=~
V^{h^0_{0k}\,,\,c} \,,\quad k\in\bbz_+ &\paran a\cr
h^0_{0k} ~=&~ h^0_{00}+\nu k\tq (\tp k +  m -\tp )
~=~ h^-_{(\tp-m, 2k\tq)} &\paran {a'}\cr
V^0_{1k} ~=&~ \(V_{1k}\)_{\vert_{n=\tq}} ~=~
\(V'_{1k}\)_{\vert_{n=\tq}} ~=~
V^{h^0_{1k}\,,\,c} \,,\quad k\in\bbz_+ &\paran b\cr
h^0_{1k} ~=&~ h^0_{10} +\nu k \tq (\tp k -  m +\tp )
~=~ h^-_{(m,(2k+1)\tq)}
&\paran {b'}\cr}$$
where ~$h^0_{0k}= h^0_{1k} = h^-_{(m,\tq)} =
{1\over 4\nu pq} [\nu^2 q^2 (\tp  - m)^2 - (p-q)^2] + \oei\mu$
~and ~$c = c^-_{p,q}\,$.
Note that $V^0_{00} = V^0_{10}\,$.

The multiplet of subtype $N^-_{21}$ describes also the situation for $\hR$
when both \redv{} and \redr\ hold, and then \rdvr\ holds.
We shall denote the corresponding multiplet by $R^-_{21}$ to stress
that it happens only for $\hR$.
The modules of this multiplet are also divided in
two infinite groups whose parametrization is obtained from
those of type $N^{1}_-$ by using \rdvr. Thus, we get:
\eqna\parar
$$\eqalignno{ \tV^0_{0k} ~=&~ \(\tV_{0k}\)_{\vert_{(m,n)=(\pot,\qot)}} ~=~
\(\tV_{1k}\)_{\vert_{(m,n)=(\pot,\qot)}} ~=~
\tV^{\th^0_{0k}\,,\,c} \,,\quad k\in\bbz_+ &\parar a\cr
\th^0_{0k} ~=&~ \th^0_{00} +2 p q k^2
~=~ h^-_{\((4k+1)\pot,\qot\)} ~=~
h^-_{\(\pot, (4k+1)\qot\)}
&\parar {a'}\cr
\tV^0_{1k} ~=&~ \(\tV'_{0k}\)_{\vert_{(m,n)=(\pot,\qot)}}
~=~  \(\tV'_{1k}\)_{\vert_{(m,n)=(\pot,\qot)}} ~=~
\tV^{\th^0_{1k}\,,\,c} \,,\quad k\in\bbz_+ &\parar b\cr
\th^0_{1k} ~=&~ \th^0_{10} +  2 p q (k+\half)^2 ~=~
h^-_{\((4k+3)\pot,\qot\)} ~=~
h^-_{\(\pot, (4k+3)\qot\)}
&\parar {b'}\cr}$$
where ~$\th^0_{00} = h^-_{(\pot,\qot)} = -
{1\over 8 pq}\, (p-q)^2  + \osi ~=~ \osic ~=~ \osi\,
c^-_{p,q}\,$. We remind that the modules here are
ordinary Verma modules with highest weight vector $J_0\,v_0\,$.
Note that the tildes are omitted in Table~1 since we use
the same diagram as for $N^-_{21}\,$.

The (generalized) Verma modules of subtype $N^{21}_+$ for fixed parameters
$p,q,m$ also form a multiplet represented by the corresponding
diagram of Table~1. The modules of this multiplet are also divided in
two groups which we denote by $V^{+0}_{0k}\,$,
$V^{+0}_{1k}\,$,  which are given by \paran{}
with the changes ${\tilde p} \to -{\tilde p}, m\to -m$,
~$h\to h^+_{(m,\tq)} = {1\over 4\nu pq} [(p+q)^2 - \nu^2 q^2 (\tp - m)^2]
+ \oei\mu$, ~$c\to c^+_{p,q}\,$.

The multiplet $N^+_{21}$ describes also the situation for $\hR$
when both \redv{} and \redr\ hold, and then \rdvr\ holds.
The corresponding multiplet is denoted by $R^+_{21}$ to stress
that it happens only for $\hR$.
The modules of this multiplet are also  divided in
two infinite groups whose parametrization is obtained from
those of type $N^{1}_+$ by using \rdvr. Thus, we get:
\eqna\parap
$$\eqalignno{ \tV^{+0}_{0k} ~=&~
\(\tV^+_{0k}\)_{\vert_{(m,n)=(\pot,\qot)}} ~=~
\(\tV^+_{1k}\)_{\vert_{(m,n)=(\pot,\qot)}} ~=~
\tV^{\th^{+0}_{0k}\,,\,c}  &\parap a\cr
&\th^{+0}_{0k} ~=~ \th^{+0}_{00} - 2 p q k^2 &\parap {a'}\cr
\tV^{+0}_{1k} ~=&~ \(\tV^{'+}_{0k}\)_{\vert_{(m,n)=(\pot,\qot)}}
~=~  \(\tV^{'+}_{1k}\)_{\vert_{(m,n)=(\pot,\qot)}} ~=~
\tV^{\th^{+0}_{1k}\,,\,c}  &\parap b\cr
&\th^{+0}_{1k} ~=~ \th^{+0}_{10} - 2 p q (k+\half)^2
&\parap {b'}\cr}$$
where ~$k\in\bbz_+\,$, ~$\th^{+0}_{00} = h^+_{(\pot,\qot)} =
{1\over 8 pq}\, (p+q)^2  + \osi ~=~ \osic ~=~ \osi\,
c^+_{p,q}\,$. We remind that the modules here are
ordinary Verma modules with highest weight vector $J_0\,v_0\,$.
Note that the tildes are omitted in Table~1 since we use
the same diagram as for $N^+_{21}\,$.

We continue with subtype $N^{22}_-\,$.
For fixed parameters $p,q,\eps$ (cf. Table 2)
the (generalized) Verma modules of this subtype form a multiplet
represented by the corresponding  diagram
of Table~1. The modules of the multiplet are an infinite set
whose parametrization is obtained from those of subtype $N^{21}_-$ by
setting  ~$m=\tp$~ when $\eps=0$~:
\eqna\parw
$$\eqalignno{
V^0_{k} ~=&~ (V^0_{0k})_{\vert_{m=\tp}} ~=~
(V^0_{1k})_{\vert_{m=\tp}} ~=~  V^{h^0_k\,,\,c}
&\parw a\cr
h^0_k ~=&~ h^0_0+\nu \tq \tp k^2
~=~ h^-_{(\tp,(2k+1)\tq)} }$$
where ~$k\in\bbz_+\,$,
~$h^0_0 = h^-_{(\tp,\tq)} =
-{1\over 4\nu pq} (p-q)^2 + \oei\mu$~ and ~$c = c^-_{p,q}\,$,
or setting ~$m=0$~ when $\eps=1$~:
$$\eqalignno{
V^1_{k} ~=&~ (V^0_{1k})_{\vert_{m=0}} ~=~
(V^0_{0,k+1})_{\vert_{m=0}} ~=~
V^{h^1_k\,,\,c} &\parw b\cr
h^1_k ~=&~ h^1_0+\nu \tp \tq k(k +1)
~=~ h^-_{(\tp, 2(k+1)\tq)}
}$$
where ~$k\in\bbz_+\,$, ~$h^1_0 = h^-_{(0,\tq)} =
{1\over 4\nu pq} [\nu^2 p^2 \tq^2 - (p-q)^2] + \oei\mu $~
and ~$c = c^-_{p,q}\,$.
Note that the case \parw{a,a'} it is necessary that $\tp-\tq \in \bbz+\mu$,
which excludes the $\hR$ case since $\tp-\tq \in \bbz$ in all cases.
In the case \parw{b,b'} it is necessary that  $\tp \in \bbz+\mu$,
which excludes  the $\hR$ case if $pq\in 2\bbn$ and
excludes  the $\hS$ case if $pq\in 2\bbn +1$.

The (generalized) Verma modules of subtype $N^{22}_+$ for
fixed parameters $p,q,\eps$
also form a multiplet represented by the corresponding
diagram of Table~1. The modules of this multiplet also
form an infinite set we denote by $V^{+\eps}_{k}\,$,
whose parametrization is obtained from \parw,\parw, resp.,
for $\eps=0,1$, resp., with the changes
~$h\to h^+_{(\tp,\tq)} = {1\over 4\nu pq} (p+q)^2
+ \oei\mu$, ~$h\to h^+_{(0,\tq)} = {1\over 4\nu pq}
[(p+q)^2 - \nu^2 q^2 \tp^2] + \oei\mu$, resp.,
and ~${\tilde p} \to -{\tilde p}$, $c\to c^+_{p,q}\,$.

We continue with subtype $N^{23}_-\,$.
For fixed parameter $\eps=0,1$ (cf. Table 2)
the (generalized) Verma modules of this subtype form a multiplet
represented by the corresponding  diagram
of Table~1. The (generalized) Verma modules of this subtype
form an infinite set
whose parametrization is obtained from those of subtype $N^{22}_-$ by
setting  ~$p=q=1$; then $\tp=\tq =\nnu\,$;~
~$c = c^-_{1,1} =1$.  We have for ~$\eps=0$~:
\eqn\parwa{ V^0_{k} ~=~ V^{h + \nnu k^2,1} }
where ~$k\in\bbz_+\,$, ~$h= h^-_{(\half,\half)} =  \oei\mu = 0$.
The last equality is because this case is possible only for $\hW,\hS$.
 ~For ~$\eps=1$~ we have:
\eqn\paroa{ V^1_{k} ~=~ V^{h+  \nnu  k(k +1),1} }
where ~$k\in\bbz_+\,$, ~$h= h^-_{(0,\half)} =
\qtrn  + \oei\mu$.
This case is possible only for $\hW,\hR$.

Finally, we consider subtype $N^{23}_+\,$.
For fixed parameter $\eps=0,1$ (cf. Table 2)
the (generalized) Verma modules of this subtype form a multiplet
represented by the corresponding diagram of Table~1.
The (generalized) Verma modules of this subtype form an infinite set
whose parametrization is obtained from those of subtype $N^{22}_+$ by
setting  ~$p=q=1$; then $\tp=\tq =\nnu\,$;~
~$c = c^+_{1,1} =c_0$.  We have for ~$\eps=0$~:
\eqn\parwa{ V^{+0}_{k} ~=~ V^{h - \nnu k^2,1} }
where ~$k\in\bbz_+\,$, ~$h= h^+_{(\half,\half)} = \nnu   + \oei\mu
= \nnu$. The last equality is because this case is possible only for
$\hW,\hS$.  ~For ~$\eps=1$~ we have:
\eqn\paroa{ V^1_{k} ~=~ V^{h -  \nnu  k(k +1),1} }
where ~$k\in\bbz_+\,$, ~$h= h^+_{(0,\half)} =
{3\over 4\nu } + \oei\mu$.
This case is possible only for $\hW,\hR$.

\newsec{Characters of (generalized) highest weight modules}

\nt
We recall the weight space decomposition of $V^{h,c}$
\eqn\wsd{   V^{h,c} ~=~ \mathop{\oplus}\limits_j V^{h,c}_j, \ \
j\in \bbz_+\ \  {\rm for} \ \  \h{W}, \h{R}, \ \
j\in \half\bbz_+\ \ \ {\rm for} \ \h{S} ~, }
where $V^{h,c}_j$ are eigenspaces of $L_0$
\eqn\eig{ V^{h,c}_j ~=~ \{v\in V^{h,c}\vert L_0v = (h+j)v\}
\cong U(\h{Q}_-)_j
{\tilde v}_0 ,}
where the last equality follows from
\eqn\deco{   U(\h{Q}_-) = \mathop{\oplus}\limits_j U(\h{Q}_-)_j}
with the range of $j$ as in (16). For $\h{R}$ we have also
\eqn\rrr{\eqalign{   &V^{h,c}_j = V^{h,c}_{j,1} \oplus V^{h,c}_{j,2} ~,\cr
& V^{h,c}_{j,1} = U(\h{Q}_-)_j\, J_0v_0, \ \
V^{h,c}_{j,2} = U(\h{Q}_-)_j\, v_0 ~.}}

The character of $V^{h,c}$ is defined (cf. \Kac,\FQS) as
\eqn\charv{  ch \ V^{h,c}(t) ~=~ \sum_j\, (\dim V^{h,c}_j)\, t^{h+j}
= t^h\, \sum_j p(j)\, t^j = t^h
\psi(t) ~, }
where $p(j)$ is the partition function ($p(j) = \#$ of ways $j$ can be
represented as the sum of positive integers (and half--integers for $\h{S}$);
$p(0) \equiv 1$), while $\psi(t)$ is given by \Kac,\FQS:
\eqn\ips{   \psi(t) =\cases{
\prod_{k\in\bbn}\, (1-t^k)^{-1} & for $\h{W}$ \cr
\prod_{k\in\bbn}\, (1+t^{k-1/2})/(1-t^k) & for $\h{S}$\cr
\prod_{k\in \bbn}\, 2(1+t^k)/(1-t^k) & for $\h{R}$ \cr}}
The factor of\ 2\ for $\h{R}$ appears because of the relation
\eqn\ror{   \dim V^{h,c}_j = (1 + 2\mu)\dim U(\h{Q}_-)_j\ .}
For $\h{R}$ and $h = c/16$ we have for
${\tilde V} = {\tilde V}^{c/16,c}\equiv U(\tilde R_-)\, J_0 v_0$
\eqn\chrr{    ch \ {\tilde V}(t) =
t^{c/16}\prod_{k\in\bbn}(1+t^k)/(1-t^k) ~.}

We present now the character formulae for the irreducible
GHWM ~$L_\L$~ in the cases when $L_\L \neq V^\L$
(for ${\hW}$ cf. \FFc, \RC, \Dosr, (for partial results see \Kac,
\Kaca, \RCW, \KW, \IZ); for
$\h{S},\h{R}$ cf. \Dosr,
(for partial results see \Kac, \KW, \GKO)).

In the $N^0$ case the embedded $V^{h+\nu mn,c}$ is irreducible while for
$L^{h,c}$ using the results of \RC\ one can obtain
\eqn\chno{  ch \, L^{h,c} ~=~ ch \, V^{h,c} - ch \, V^{h+\nu mn,c}
  ~=~ (1-t^{\nu mn})\ ch \, V^{h,c} ~.}

In the $N^1_-$ case let us denote by
$L_{0k}, L_{1k}, L'_{0k}, L'_{1k}$
the irreducible factor--modules of $V_{\ell k}, V'_{\ell k}\,$,
resp. Then we have (\FFc, \RC, \Dosr, formula (22)):
\eqna\chnon
$$\eqalignno{  &ch \, L_{\ell k} ~=~ ch \ V_{\ell k} + \sum_{j>k}
(ch\ V_{0j} + ch \ V_{1j}) - \sum_{j\geq k} (ch \
V'_{0j} + ch \ V'_{1j}), &\chnon a\cr
&   ch \, L'_{\ell k} ~=~ ch \, V'_{\ell k} + \sum_{j>k}(ch \ V'_{0j}
+ ch \ V'_{1j} - ch \ V_{0j} - ch \ V_{1j}), &\chnon b\cr}$$
where in both formulae $\ell = 0,1$, ~$k\in\bbz_+\,$.

In the $N^1_+$ case we denote by
$L^+_{\ell k}, L^{'+}_{\ell k}, \ (\ell = 0,1)$, the
irreducible factor--modules of
$V^+_{\ell k}, V'^+_{\ell k}$ resp.
Then we have (\FFc, \RC, \Dosr, formula (35)):
\eqna\chpp
$$\eqalignno{ch \, L^+_{\ell k} ~=&~ ch \, V^+_{\ell k} + \sum^{k-1}_{j=1}
(ch \, V_{0j} + ch \, V_{1j}) + ch \, L^+_{00} - \sum^{k-1}_{j=0}
(ch \, V'_{0j} + ch \, V'_{1j}), \ \
k> 0 \quad\quad\quad &\chpp a\cr
ch \, L'^+_{\ell k} ~=&~ ch \, V'^+_{\ell k} + \sum^{k-1}_{j = 0}
(ch \, V'_{0j}+ ch \, V'_{1j}) - ch \, L^+_{00} -\sum^k_{j=1}
(ch \, V_{0j} + ch \, V_{1j}), \ \ k \geq 0 \quad\quad\quad &\chpp b\cr }$$
where in both formulae $\ell = 0,1$;
($ch \, L^+_{00} = ch \, L^+_{10} = ch \, V^+_{00} = ch \, V^+_{10}$,
since $V^+_{00} = V^+_{10}$ is irreducible).

In the $N^{21}_\pm$ cases let us denote by
$L^0_{\ell k}\,$, $L^{+0}_{\ell k}$ the irreducible
factor module of $V^0_{\ell k}\,$, $V^{+0}_{\ell k}\,$, resp.
Then we have (\FFc, \RC, \Dosr, formulae (28), (29)):
\eqn\chto{\eqalign{ ch \, L^0_{0k} ~=&~
ch \, V^0_{0k} - ch \, V^0_{1k}, \ \  k > 0, \cr
ch \, L^0_{1k} ~=&~
ch \, V^0_{1k} - ch \, V^0_{0,k+1}, \ \  k \geq 0, \cr }}
and (\FFc, \RC, \Dosr, formulae (42), (43)):
\eqn\chtop{\eqalign{ch \, L^{+0}_{0k} ~=&~
ch \, V^{+0}_{0k} - ch\, V^+_{1,k-1}, \ \ k > 0 ,\cr
ch \, L^{+0}_{1k} ~=&~
ch \, V^{+0}_{1k} - ch\, V^+_{0k}, \ \ k > 0 ,\cr }}
($ch \, L^{+0}_{00} = ch \, L^{+0}_{10} = ch \,
V^{+0}_{00} = ch \, V^{+0}_{10}$,
since $V^{+0}_{00} = V^{+0}_{10}$ is irreducible.)
These formulae are valid also for the $R^{21}_\pm$ cases
but all generalized Verma modules $V^0_{\ell k}\,$, $V^{+0}_{\ell k}\,$
should be replaced with the ordinary
Verma modules $\tV^0_{\ell k}\,$, $\tV^{+0}_{\ell k}\,$,
(with highest weight vector $J_0\,v_0\,$ as discussed above).

In the $N^{22}_\pm\,,N^{23}_\pm\,,$ cases let us denote by
$L^\eps_{k}\,$, $L^{+\eps}_{k}$ the irreducible
factor module of $V^\eps_{k}\,$, $V^{+\eps}_{k}\,$, resp.
Then we have (\FFc, \RC, \Dosr, formula (31)):
\eqn\chty{ch \, L^\eps_{k} ~=~
ch \, V^\eps_{k} - ch \, V^\eps_{k+1}, \ \  k \geq 0, }
and (\FFc, \RC, \Dosr, formulae (44), (45)):
\eqn\chty{ch \, L^{+\eps}_{k} ~=~
ch \, V^{+\eps}_{k} - ch \, V^{+\eps}_{k-1}, \ \  k > 0. }
($ch\,L^{+\eps}_{0} ~=~ ch \, V^{+\eps}_{0}$ since $V^{+\eps}_{0}$
is irreducible.)

\newsec{Unitarity}

\nt
The irreducible factor-modules
~$L_\L ~=~ L^{h,c} ~=~ V^\L/I^\L$~ are unitarizable \FQS,\GKO,\Lan\
if ~{\it either}~ $h\geq 0$, $c\geq 1$,
~{\it or}~ when
\eqn\ccc{h  ~=~  {1\over 4\nu p(p+\nu)}
[\nu^2(pn - m(p+\nu))^2 - \nu^2] + \oei\mu\,, \quad c ~=~ 1 - {\textstyle{2(2+\nu)\over p(p+\nu)}}
\,, \quad p=2,3,\ldots   }
~{\it or}~ when
\eqn\cccc{h  ~=~ h^-_{(\pot,\pott)} ~=~ \osi c^-_{p,p+1} ~=~ \osi \(1- {\textstyle{2\over p(p+1)}}\)
\,, \quad p=1,2,\ldots \, \qquad {\rm for}~~~ \hR\ .}
The cases \ccc\ are
the modules $L_{00} \equiv L_{10}$ (parametrized by $(m,n)$ as in Table 2)
which form the $c<1$ series of unitarizable HWM over
$\widehat W, \FQS,\Lan, \widehat S, \widehat R$ \FQS,\GKO\ inside the so-called
Kac table \BPZ. The case \cccc\ is the module
  $L^0_{00} \equiv L^0_{10}\,$ on the border of the Kac table.

In the cases \ccc\ for $p=2$ one gets the trivial case $c=0$.
The cases $p=3,4,5,6$ for the Virasoro algebra ($\nu=1$),
i.e., ~$c ~=~ 1/2,\ 7/10,\ 4/5,\ 5/6\,$, resp.,
correspond to the Ising model, tri-critical Ising model,
3-state Potts model, tri-critical 3-state Potts model, resp., \FQS.
Note that there is only one value of the central charge
which is common for the Virasoro and super-Virasoro algebras.
This is the case $c=7/10$ for Virasoro ($p=4$) and $c=7/15$
for the  super-Virasoro algebras ($p=3$), taking into account that
~$c_{\hW} ~=~ \thf\,c_{\hS,\hR}\,$ (compare \vir{a} and \virn{a}).

\parskip=0pt
\listrefs

\fig{}{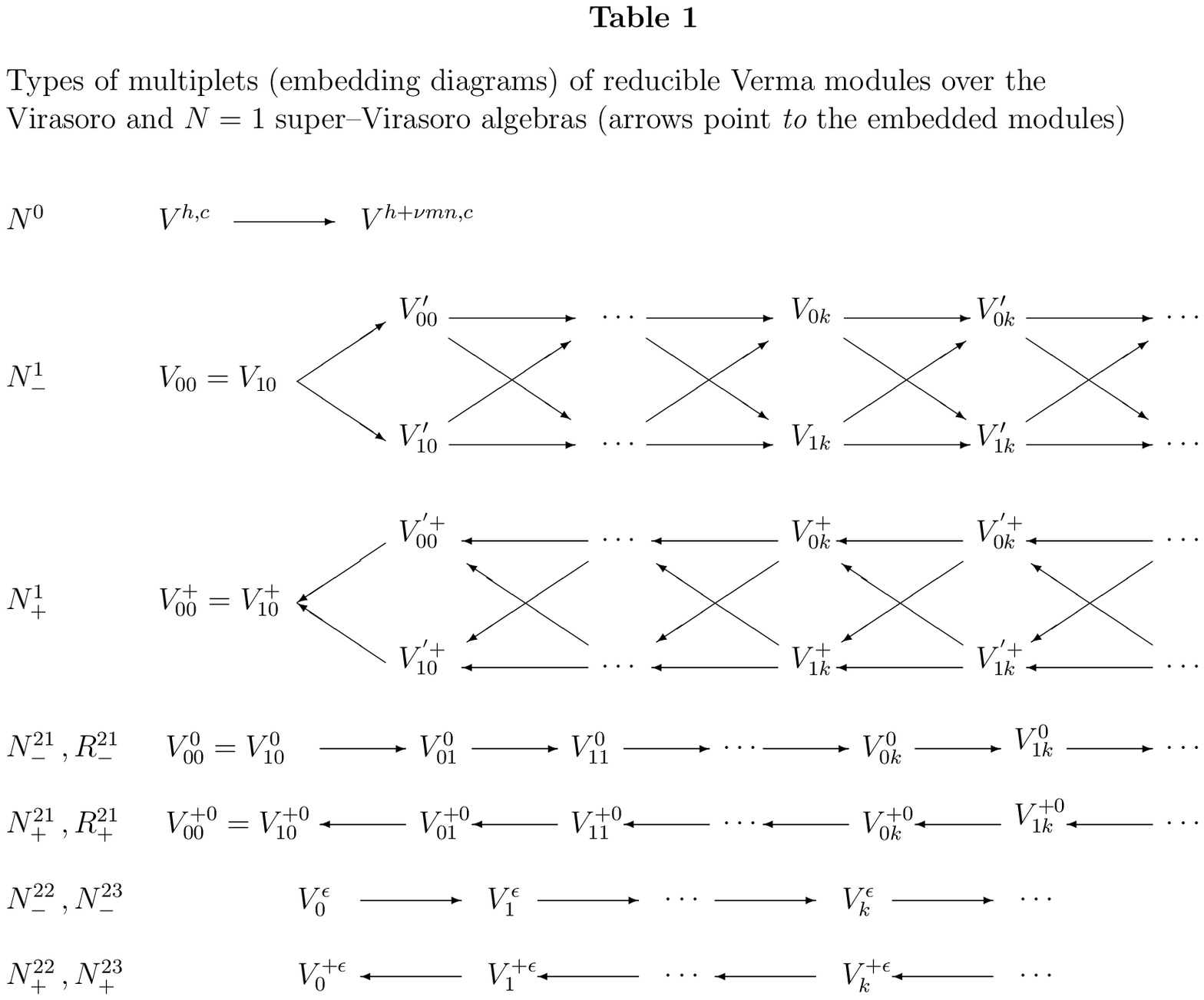}{13cm}

\fig{}{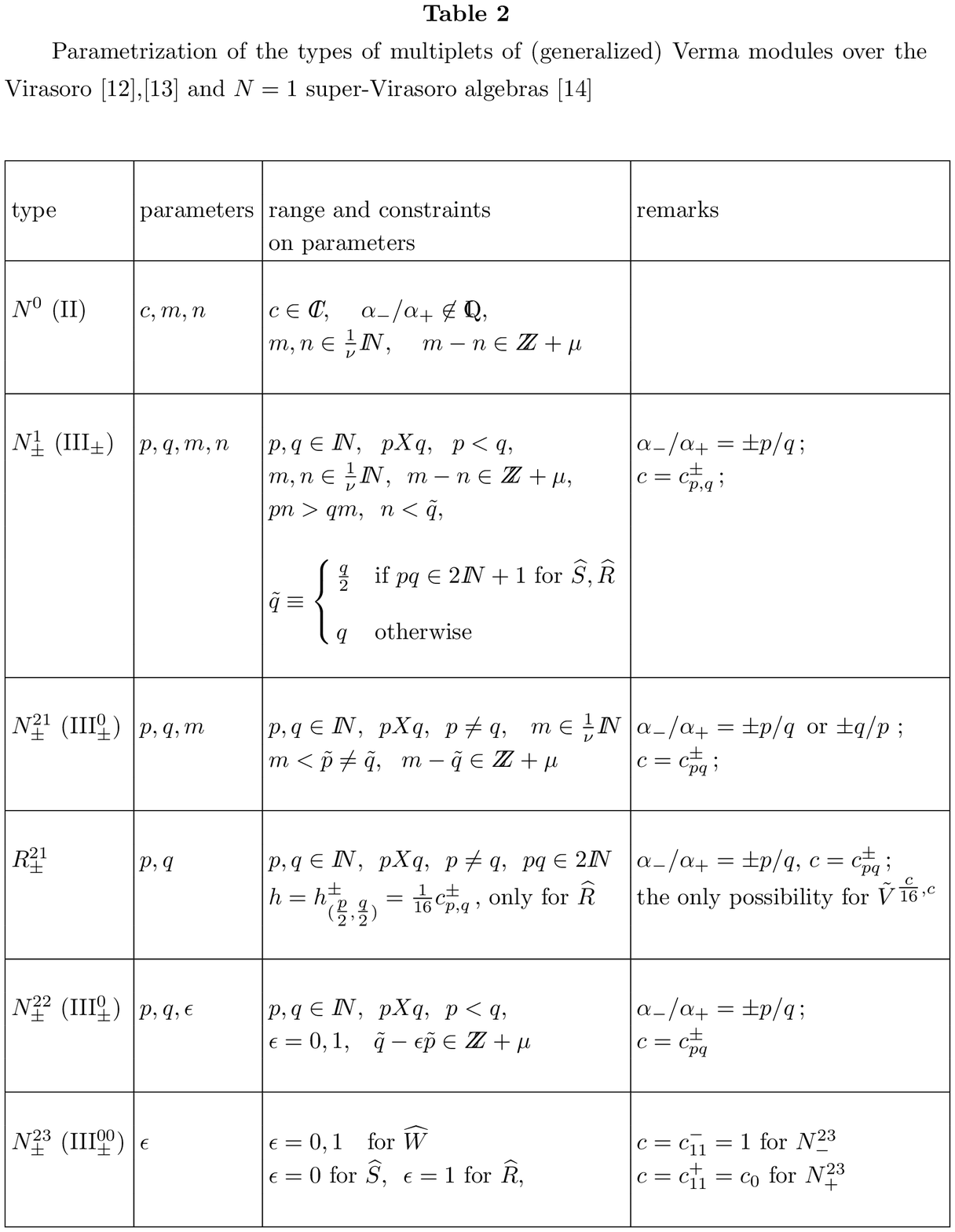}{13cm}

\end